\documentclass[10pt,letterpaper,twocolumn]{article} 

\usepackage{ol2}
\usepackage{graphicx}
\usepackage{hyperref}
\usepackage{amsmath}

\begin{document}

\twocolumn[ 

\title{Fermionic out-of-plane structure of polarization singularities}

\author{Mark R Dennis}

\address{H H Wills Physics Laboratory, University of Bristol, Tyndall Avenue, Bristol BS8 1TL, UK\\
mark.dennis@physics.org}

\begin{abstract}
A new classification of circular polarization C points in three-dimensional polarization ellipse fields is proposed.
The classification type depends on the out-of-plane variation of the polarization ellipse axis, in particular, whether the ellipse axes are in the plane of circular polarization one or three times.
A minimal set of parameters for this classification are derived, and discussed in the context of the familiar in-plane C point classification into lemon, star, and monstar types.
This new geometric classification is related to the M\"obius index of polarization singularities recently introduced by Freund.
\end{abstract} 


] 

A goal of modern optics is a complete understanding of the fine interference structure of classical monochromatic fields.
At its most general, a nonparaxial, complex vector light field in three dimensions has elliptic polarization at each point, in part characterized by the ellipse major axis, whose direction in three dimensions varies with position.
The ellipse axis is a director (represented by $\pm$ a vector), and this director direction is undefined and topologically singular when the polarization is circular, which occurs along space curves in three dimensions called `C lines' \cite{nh:1987structure}.
These C lines are topological defects in the smooth complex vector fields describing the light field, and on a loop around the C line, the director undergoes a rotation through $\pi,$ fermionically exchanging its two sides (i.e.~switching sign after a rotation by $2\pi$).
These polarization singularities have been the subject of much study (reviewed in Refs.~\onlinecite{nye:1999natural,dop:2009review}), and 2D properties have been measured \cite{des:2004measurement,fodp:2008polarization}, although measurement of 3D fields is still a challenge at optical frequencies.

Topological defects have topological and geometric properties, and in Ref.~\onlinecite{nh:1987structure}, Nye and Hajnal identified several of these for points on C lines looking at the nearby ellipses projected into the plane of the circular polarization like their 2D counterparts \cite{nye:1983cline}: e.g.~the $\pm1/2$ topological index of the director's rotation about the C point, and the `L classification' \cite{bh:1977b60} according to which the projected axis becomes purely tangential to a circle around the C point once or three times, denoted L1 or L3.
There are three types of two-dimensional C point, `lemon' ($+1/2$ L1), `star' ($-1/2$ L3) and the transitional `monstar' ($+1/2$ L3), and these are determined by two parameters describing the nonuniform rate of change of the director around the C point\cite{dennis:2008monstardom}.

However, such properties are purely 2-dimensional, and the ellipses close to a C point also vary out of the plane of circular polarization.
As a way of understanding this, Freund \cite{freund:2010mobiusc} recently proposed polarization `M\"obius bands' around C points: on a loop enclosing the C point, the ellipse axis generically winds $\pm1/2$ or $\pm3/2$ times around the loop, this twist number being half-integer due to the fermionic C point index.
Such M\"obius band geometry is, however, very complicated, depending on both the in-plane and out-of-plane C point geometry and no mathematical analysis has been made of them.

\begin{figure}
   \centerline{\includegraphics[width=5.3cm]{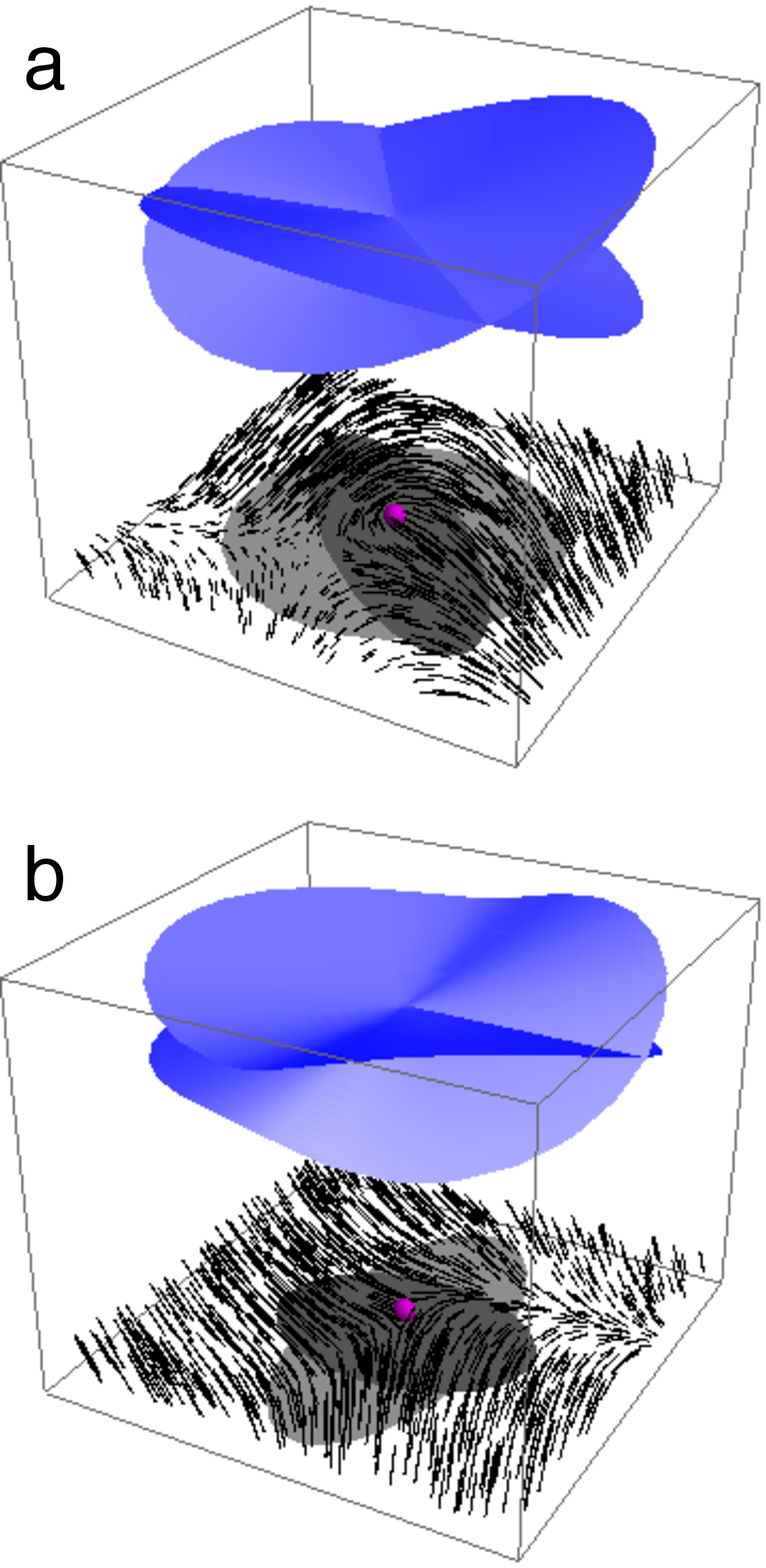}}
   \caption{(Color online)   
   Representation of polarization axes close to C points. 
   (a) $\boldsymbol{E} = (\sqrt{2}+x+\mathrm{i} y, \mathrm{i}\sqrt{2}-\mathrm{i} x + y,(-1.40-.18\mathrm{i})x+(-1.40+.18\mathrm{i})y).$ (b) $\boldsymbol{E} = (1. - 0.98 x + 0.20 \mathrm{i} y, \mathrm{i} + 0.98 \mathrm{i} x + 0.20 y,  0.63 (x -  2 \mathrm{i} y)).$
   In each frame, the black lines represent the ellipse axes around the C point (sphere), in its plane of polarization.
   The lighter (darker) gray ellipse is the anisotropy ellipse for $E_-$ ($E_z$), and the blue surface above represents $\pm a_z$ around the C point, with either three zeros (a) or one zero (b), resembling a 2-sheet Riemann surface. 
   }
   \label{fig:sheets}
\end{figure}

Here, I will describe an alternative, simpler geometric classification, the `O classification' of the out-of-plane variation around a C point in 3D complex vector fields: the number of directions out from the C point in which the axis lies in the plane of the C point is generically one or three.
The odd number is the signature of a fermionic multivaluedness: there are two opposite ellipse axis vectors, which must be exchanged in a circuit around the C point.
This contrasts with the `bosonic' out-of-plane variation around a typical polarization ellipse, for which there are generically two opposite directions in which the local ellipse axes lie in the the central polarization plane (as described below).
Examples of the possible fermionic out-of-plane structures are represented in Fig.~\ref{fig:sheets}.

In considering the mathematical basis of this O classification, it is natural to choose a local helical basis $\boldsymbol{E}=(E_+,E_-,E_z)$ for the complex vector field around the point of interest in three dimensions (placed at the origin), with $E_z = 0$ (i.e.~the polarization ellipse is in the $xy$ plane).
In this basis, the helical unit vectors are $\boldsymbol{e}_{\pm} \equiv \frac{1}{\sqrt{2}}(1, \pm \mathrm{i},0)$ with respect to cartesian coordinates, and $|E_+|\ge|E_-|$ without loss of generality.
Since $E_z = 0$ at the origin, there is a complex phase singularity in the local $z$-component, and at an ordinary ellipse, $E_+,E_-$ are nonzero constants.
At a C point, $E_- = 0$ at the origin as well, leaving only $E_+$ nonzero, and so in this plane the C point appears right-circular.

The choice of a helical basis makes the in-plane C point classification straightforward.
The projected, in-plane ellipse axis direction is given by half the argument of $\sigma = S_1 + \mathrm{i} S_2 = E_+^* E_-$ where $S_1, S_2$ are two-dimensional Stokes parameters in the plane\cite{freund:poincare,dennis:2002polarization}: at a C point, this in-plane axis angle singularity is directly related to the phase singularity in $E_-.$
It is further convenient to choose local coordinates and phase such that $\nabla E_- = (u_x, \mathrm{i} v_y,0)$ with $|v_y| \ge u_x \ge 0,$ so the sign of $v_y$ determines the C point index, and the anisotropy ellipse\cite{ss:parameterization,bd:2000b321,roux:coupling} of the phase singularity in $E_-$ (and thus the C point) is aligned with the $xy$-axes. 
With this choice, $-\frac{1}{2}\arg E_+ \equiv \beta$ is the in-plane axis orientation angle along the local $x$-direction, and in Ref.~\onlinecite{dennis:2008monstardom} it was shown that the index and L type depends only on the angle $\beta$ and the C point isotropy, parametrized by $\Upsilon \equiv 2u_x v_y/(u_x^2 + v_y^2).$

The complex quantity $\boldsymbol{E}\cdot\boldsymbol{E} \approx E_+ E_-$ close to a C point is related to $\sigma.$ 
$|\boldsymbol{E}\cdot\boldsymbol{E}| = |\sigma|$ is a measure of the polarization ellipticity, and $\frac{1}{2}\arg \boldsymbol{E}\cdot\boldsymbol{E}$ is the vibration phase, determining where $\mathrm{Re}\boldsymbol{E}$ is around the ellipse \cite{nye:1983cline,nye:1999natural,bd:2001b324}, which is zero when the real part is along the major axis of the ellipse, and undefined for circular polarization.
With the coordinate and parameter choices above, at a C point, $\frac{1}{2}\arg \boldsymbol{E}\cdot\boldsymbol{E} = -\beta$ with respect to the $+x$-direction.

The director field of ellipse axes is conveniently represented by the smooth field via the vibration phase \cite{berry:2004,dennis:2002polarization}
\begin{equation}
   \boldsymbol{a} = \pm \mathrm{Re}\sqrt{\boldsymbol{E}^*\cdot\boldsymbol{E}^*}\boldsymbol{E}.
   \label{eq:afield}
\end{equation}
At a C point, $\boldsymbol{a} = 0,$ and in a loop around the point, $\pm\to\mp,$ revealing the singularity's fermionic nature.
The O classification is determined by the $z$-component $a_z$ in a neighborhood of the C point.

Close to a typical point of elliptical polarization, placed at the origin with the coordinates chosen above,
\begin{equation}
   a_z \approx \pm \mathrm{Re}\sqrt{E_+^* E_-^*}(x,y,0)\cdot\nabla E_z,
   \label{eq:typical}
\end{equation}
where $E_+^* E_-^*$ is a nonzero complex constant evaluated at the origin.
Therefore $a_z = 0$ locally along the two lines perpendicular to $\pm\mathrm{Re}\sqrt{E_+^* E_-^*}\nabla E_z,$ and $\pm a_z$ is represented by a pair of tipped planes through the C point.

\begin{figure}
   \centerline{\includegraphics[width=7cm]{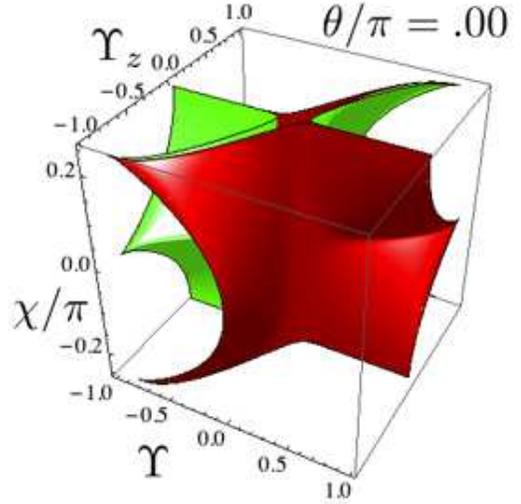}}
   \caption{(Color online)   
   Surface $D_{\mathrm{O}} = 0$ for alignment $\theta = 0,$ in $(\Upsilon, \Upsilon_z, \chi)$-space, where $D_{\mathrm{O}}>0$ on the red side.
   The surface is periodic in $\chi,$ and in the four faces where $\Upsilon, \Upsilon_z = \pm 1,$ the line $D_{\mathrm{O}}=0$ resembles the L classification locus \cite{dennis:2008monstardom}.
   }
   \label{fig:Oclass}
\end{figure}

At a C point, both $E_-$ and $E_z$ are 0, and in this case, the out-of-plane component of the axis, in the coordinates chosen above, is
\begin{equation}
   a_z \approx \pm \mathrm{Re}|E_+|^{1/2}\mathrm{e}^{\mathrm{i}\beta}\sqrt{u_x x - \mathrm{i}v_y y}(x,y,0)\cdot\nabla E_z.
   \label{eq:azc}
\end{equation}
The variation of $a_z$ therefore depends on the shapes of the two phase singularities, one in $E_-$ (which is square rooted), and the other in $E_z.$
The structure may be understood by writing the complex vector $\nabla E_z = \mathrm{e}^{2\mathrm{i}\nu}(\cos{\theta}U_X -\mathrm{i} \sin\theta V_Y, \sin{\theta} U_X + \mathrm{i} \cos\theta V_Y),$ with $|V_Y| \ge U_X \ge 0.$
This can be interpreted as the anisotropy ellipse of the nodal point in $E_z$ having axes proportional to $U_X, V_Y$ (in the $X,Y$ direction), making an alignment angle $\theta$ with the $x,y$ directions defined by the axes of the anisotropy ellipse of the node in $E_-.$
The isotropy parameter of this $E_z$ anisotropy ellipse is written $\Upsilon_z = 2U_X V_Y/(U_X^2+V_Y^2),$ similar to $\Upsilon$ defined above for the in-plane isotropy.
The angle $\nu$ is the vibration phase associated with the complex vector $\nabla E_z:$ the phase factor multiplying the complex $E_z$ to find the real $a_z$ depends on a combination of $\nu,$ the vibration phase of the in-plane pattern $\beta,$ and the alignment angle $\theta.$

The O classification of C points, that is, whether $a_z = 0$ once or thrice on a small loop around the C point, is determined in a similar way to the L classification\cite{bh:1977b60,dennis:2002polarization}.
The loop $(x,y)=\delta(\cos t, \sin t \, |v_y|/u_x),$ corresponding to a circuit on the anisotropy ellipse itself, is a more convenient choice than the circle.
Setting $a_z = 0$ from Eq.~(\ref{eq:azc}) on this loop and rearranging gives a cubic polynomial equation in $\tan(t/2),$ with real coefficients depending on the in-plane isotropy $\Upsilon,$ the out-of-plane isotropy $\Upsilon_z$ the anisotropy ellipse alignment angle $\theta,$ and the phase $\chi \equiv \beta + 2\nu,$ a combination of the vibration phases of the two anisotropy ellipses.
This polynomial has one or three roots, corresponding to one or three zeros of $a_z$ on the loop (O1, O3 respectively), depending on the sign of the discriminant of this cubic, given by
\begin{align}
 & D_{\mathrm{O}}  =  2 (3 - \Upsilon^2 - 14 \Upsilon \Upsilon_z - \Upsilon_z^2 + 12 \Upsilon^2 \Upsilon_z^2) \nonumber \\
   & \quad - (1 + \Upsilon) \sqrt{1 - \Upsilon^2} (1 + \Upsilon_z)^2 \cos4 (\chi - \theta) \nonumber \\
   & \quad + 2 (2 - \Upsilon) (1 + \Upsilon) (1 + \Upsilon_z) \sqrt{1 - \Upsilon_z^2} \cos2(2 \chi - \theta) \nonumber \\
   & \quad - 4 \sqrt{1 - \Upsilon^2} (2 - 7 \Upsilon \Upsilon_z) \sqrt{1 - \Upsilon_z^2} \cos 2 \theta \nonumber \\
   & \quad + 2  (1 - \Upsilon^2)  (1 - \Upsilon_z^2) \cos4 \theta - 6 \sqrt{1 - \Upsilon^2}  (1 - \Upsilon_z^2) \cos 4 \chi \nonumber \\
   & \quad + 2 (1 - \Upsilon) (2 + \Upsilon) (1 - \Upsilon_z) \sqrt{1 - \Upsilon_z^2} \cos2(2 \chi + \theta) \nonumber \\
   & \quad - (1 - \Upsilon) \sqrt{1 - \Upsilon^2} (1 - \Upsilon_z)^2 \cos4(\chi + \theta),
   \label{eq:DO}
\end{align}
with $D_{\mathrm{O}} > 0$ for O3 ($< 0$ for O1). 
Equation (\ref{eq:DO}) is the main result of this Letter.

The out-of-plane classification is determined by Equation (\ref{eq:DO}), depending on the four parameters $\Upsilon, \Upsilon_z, \theta,$ and $\chi,$ which is therefore hard to visualize.
A particular slice through this four-dimensional parameter space, with $\theta = 0$ (the two anisotropy ellipses are aligned), is shown in Fig.~\ref{fig:Oclass}.
The vertical faces of the figure demonstrates a property of the set $D_{\mathrm{O}} = 0:$ when the in-plane C point is isotropic ($\Upsilon  = \pm 1,$ corresponding to a uniform lemon or star), or the out-of plane node in $E_z$ is isotropic ($\Upsilon_z = \pm 1$), then $\theta$ is not defined and the $D_{\mathrm{O}} = 0$ locus, depending on $\chi$ and $\Upsilon$ or $\Upsilon_z,$ is a cusp, similar to the corresponding $D_{\mathrm{L}}$ discriminant determining the in-plane L classification, determined by $\Upsilon$ and $\beta$ \cite{dennis:2008monstardom}.

This new geometric classification, characterizing the out-of-plane behavior of a nonparaxial C point, complements the traditional index, line and other in-plane C point properties.
Combined with the L classification, it also gives some insight into the M\"obius index \cite{freund:2010mobiusc}.
A polarization M\"obius band may be thought of as a ribbon around the C point with axis curve $\delta(\cos\phi,\sin\phi,0),$ framed by the projection of $\boldsymbol{a}$ into the normal plane of the curve at each point.
In order for this ribbon to have $\pm3/2$ twists, this projection must lie in the plane $a_z = 0$ three times, and also be perpendicular to the curve (i.e.~$(a_x,a_y)$ tangent to the circle) three times; a necessary condition is that both $D_{\mathrm{L}}$ and $D_{\mathrm{O}} > 0.$

The L and O classifications correspond to the count of special directions in the $x,y$ plane around a C point.
For the L classification, it is the number of times that $(a_z,a_y)$ is tangent to the circular loop (equivalent to the number of times it is radial).
The actual positions of these lines, in fact, determine whether the M\"obius index is $\pm3/2$ or $\pm1/2$ precisely, depending on how these two sets of three directions interlace.
If, in a circuit around the C point, the special directions alternate, then $\boldsymbol{a}$ must twist around the C point exactly 3/2 times, whereas, if not (i.e.~there is a repeat of the same type of direction) the ribbon twist sense changes direction, resulting in $\pm1/2$ a twist.
In principle, the two sets of three directions can be located precisely as roots of the corresponding cubic polynomials in $\tan\phi,$ although this method is algebraically taxing, and further information is required as well to determine the sign of M\"obius twisting.

Beyond the O classification of C points, there is an equivalent out-of-plane classification for points of linear polarization\cite{nh:1987structure,bd:2001b324}, and their twist indices\cite{freund:2010mobiusl}, and it may be extended to account for more twists of larger loops beyond first order \cite{freund:2011multitwist}. 
Furthermore, this new classification is very general, although challenging to measure in a real 3D optical field; similar kinds of topological defect, such as disclinations in liquid crystals, are likely to have similar variation (related to the Frank `twist' parameter close to the defect \cite{frank:1958lc}), exhibiting further geometric fermionic properties of topological defects in director fields.

I am grateful to Isaac Freund, Robert King and Brina \v{C}rnko for discussions.
My research is supported by the Royal Society of London.

\end{document}